\documentclass[sigconf]{acmart}

\AtBeginDocument{%
  \providecommand\BibTeX{{%
    \normalfont B\kern-0.5em{\scshape i\kern-0.25em b}\kern-0.8em\TeX}}}

\copyrightyear{2020}
\acmYear{2020}
\setcopyright{acmcopyright}\acmConference[ICCAD '20]{IEEE/ACM International Conference on Computer-Aided Design}{November 2--5, 2020}{Virtual Event, USA}
\acmBooktitle{IEEE/ACM International Conference on Computer-Aided Design (ICCAD '20), November 2--5, 2020, Virtual Event, USA}
\acmPrice{15.00}
\acmDOI{10.1145/3400302.3415789}
\acmISBN{978-1-4503-8026-3/20/11}

\usepackage{ctable}
\usepackage{multirow}
\usepackage{multicol}

\PassOptionsToPackage{bookmarks=false}{hyperref}

\begin{document}
\title[Hardware-aware NAS for 3D Cardiac Cine MRI Segmentation]{Towards Cardiac Intervention Assistance: Hardware-aware Neural Architecture Exploration for Real-Time 3D Cardiac Cine MRI Segmentation}

\author{
Dewen Zeng$^{1}$ \quad
Weiwen Jiang$^{1}$ \quad
Tianchen Wang$^{1}$ \quad
Xiaowei Xu$^{2}$ \quad
Haiyun Yuan$^{2,3}$ \quad  \quad  \quad  \quad
Meiping Huang$^{2}$ \quad
Jian Zhuang$^{2}$ \quad
Jingtong Hu$^{4}$ \quad
Yiyu Shi$^{1}$
}
\affiliation{%
\vspace{5pt}
\institution{$^{1}$ Department of Computer Science and Engineering, University of Notre Dame, Notre Dame, IN, USA, 46556}
\institution{$^{2}$ Guangdong Cardiovascular Institute, Guangdong Provincial Key Laboratory of South China Structural Heart Disease, Guangdong Provincial People's Hospital, Guangzhou, China, 510000}
\institution{$^{3}$ Department of Cardiology, Boston Children's Hospital, Harvard Medical School, Boston, MA, USA, 02115}
\institution{$^{4}$ Department of Electrical and Computer Engineering,
University of Pittsburgh, Pittsburgh, PA, USA, 15261}
\vspace{5pt}
}

\renewcommand{\shortauthors}{Dewen Zeng and Weiwen Jiang, et al.}

\begin{abstract}
Real-time cardiac magnetic resonance imaging (MRI) plays an increasingly 
important role in guiding various cardiac interventions.
In order to provide better visual assistance, the cine MRI frames need to be segmented on-the-fly to avoid noticeable visual lag. In addition, considering reliability and 
patient data privacy, the computation is preferably done on local hardware.
State-of-the-art MRI segmentation methods mostly focus on accuracy only, and 
can hardly be adopted for real-time application or on local hardware. 
In this work, we present the first hardware-aware multi-scale neural architecture search (NAS) framework for real-time 3D cardiac cine MRI segmentation.
The proposed framework incorporates a latency regularization term into the loss function to handle real-time constraints, with the 
consideration of underlying hardware.
In addition, the formulation is fully differentiable with respect to the architecture parameters, so that 
stochastic gradient descent (SGD) can be used for optimization to reduce the computation cost while maintaining optimization quality.
Experimental results on ACDC MICCAI 2017 dataset demonstrate that our hardware-aware multi-scale NAS framework can reduce the latency by up to 3.5$\times$ and satisfy the real-time constraints, while still achieving competitive segmentation accuracy, compared with the state-of-the-art NAS segmentation framework. 
\end{abstract}




\maketitle

\section{Introduction}

Magnetic Resonance Imaging (MRI), which supports real-time visualization of anatomy and cardiac tissues, has been used to assist a number of cardiac interventions such as myocardial chemoablation \cite{rogers2016transcatheter}, intracardiac catheter navigation \cite{gaspar2014three} and aortic valve replacement \cite{mcveigh2006real}. 
For better visual assistance, the cine MRI needs to be accurately segmented on-the-fly without noticeable visual lag. 
As stated in \cite{annett2014low,schaetz2017accelerated,iltis2015high}, the latency should be no more than 50ms and the throughput should at least match the MRI reconstruction rate of 22 frames per second (FPS). This is a quite challenging 
problem, considering the facts that 
the images are all 3D so computationally intensive, 
that the cardiac border always has ambiguity and that large variations always exist among target objects from different patients \cite{zheng20183}.
Moreover, considering reliability and patient data privacy, the computation is preferably done on local hardware. 
In consequence, how to guarantee real-time performance and hardware efficiency while maintaining high accuracy is the main challenge in segmentation for cardiac intervention assistance. This work aims to address this challenge.  

Machine learning has shown great potential in segmenting medical images \cite{ding2020uncertainty,xu2019whole,xu2018quantization}.
Existing works mostly handcrafted sophisticated neural network structures or frameworks to improve the cardiac MRI segmentation accuracy \cite{xu2018quantization,liu2019machine,xu2019whole,ding2020uncertainty,bian2020bunet,dong2020deu,wang2020ica,isensee2017automatic,zotti2018convolutional}.
Recently, we have witnessed great success of neural architecture search (NAS), which can identify more accurate and efficient neural architectures than human-invented ones for medical image segmentation applications \cite{kim2019scalable, dong2019neural, mortazi2018automatically, weng2019unet, yan2020ms}.
Most of them, however, only focus on maximizing model accuracy and do not take latency or throughput into consideration. They can neither meet the throughput and latency requirement for real-time applications nor guarantee that the model can be accommodated onto local hardware.


Recently hardware-aware NAS were proposed \cite{wu2019fbnet,cai2018proxylessnas,jiang2019accuracy,jiang2020standing,yang2020co,jiang2020hardware,jiang2020standing,jiang2020device}, which jointly identify the best architecture and hardware designs to maximize the network accuracy and hardware efficiency.
Existing hardware-aware NAS methods mainly target GPUs or Field Programmable Gate Arrays (FPGAs).
Usually a latency estimation term is added to the loss function as a reward for reinforcement learning (RL) \cite{jiang2019accuracy} or additional penalty for differentiable NAS \cite{wu2019fbnet}.
However, they mainly focus on the image classification tasks and stick to a fixed backbone architecture.
Simply applying the same strategy to image segmentation tasks may not work as the search space is quite different.
In addition, medical images typically have much higher resolutions and as a result, employing transfer learning based on existing models for image classification may not work well \cite{chen2018searching}.

To address the above challenges, in this work we propose a hardware-aware neural architecture exploration framework for real-time 3D cardiac Cine MRI segmentation, by using a multi-scale neural architecture search method MS-NAS for medical image segmentation \cite{yan2020ms}.
We first extend MS-NAS, which only takes 2D images, to process 3D cine MRI images.
Then a differentiable neural architecture search method is used to enable hardware-awareness for real-time 3D MRI segmentation tasks.
In order to estimate the latency of architecture, the latency of each operator is profiled into a lookup table. Then a layer-wise latency estimation scheme is used to quickly estimate the latency of the multi-scale supernet.
By using the Gumbel Softmax technique, the estimated network latency is fully differentiable with respect to the architecture parameters.
Therefore, conventional gradient-based optimization such as stochastic gradient descent (SGD) can be directly used for efficient optimization.

The main contribution of this work is listed as follows:
\begin{itemize}
\item We propose a 3D multi-scale neural architecture search framework for real-time 3D cardiac cine MRI segmentation tasks. The framework includes both cell level search and architecture backbone search to help identify optimal architecture for 3D MRI segmentation.
\item We explicitly incorporate a hardware latency term into the loss function and jointly optimize accuracy and hardware latency. We formulate our work as an over-parameterized differentiable mathematical problem so that SGD optimization can be used.
\item We demonstrate that the proposed 3D multi-scale NAS framework can identify neural architecture with 39ms latency and 25.6FPS throughput on ACDC 2017 dataset, satisfying real-time constraints. At the same time, comparable segmentation accuracy is achieved.  
\end{itemize}

The remainder of the paper is organized as follows. Section 2 reviews related background and motivation. Section 3 introduces the NAS search space, hardware-aware loss function, and optimization strategy. Experimental results are shown in Section 4 and concluding remarks are given in Section 5.

\begin{figure*}[t]
\centering
\includegraphics[width=1.9 \columnwidth]{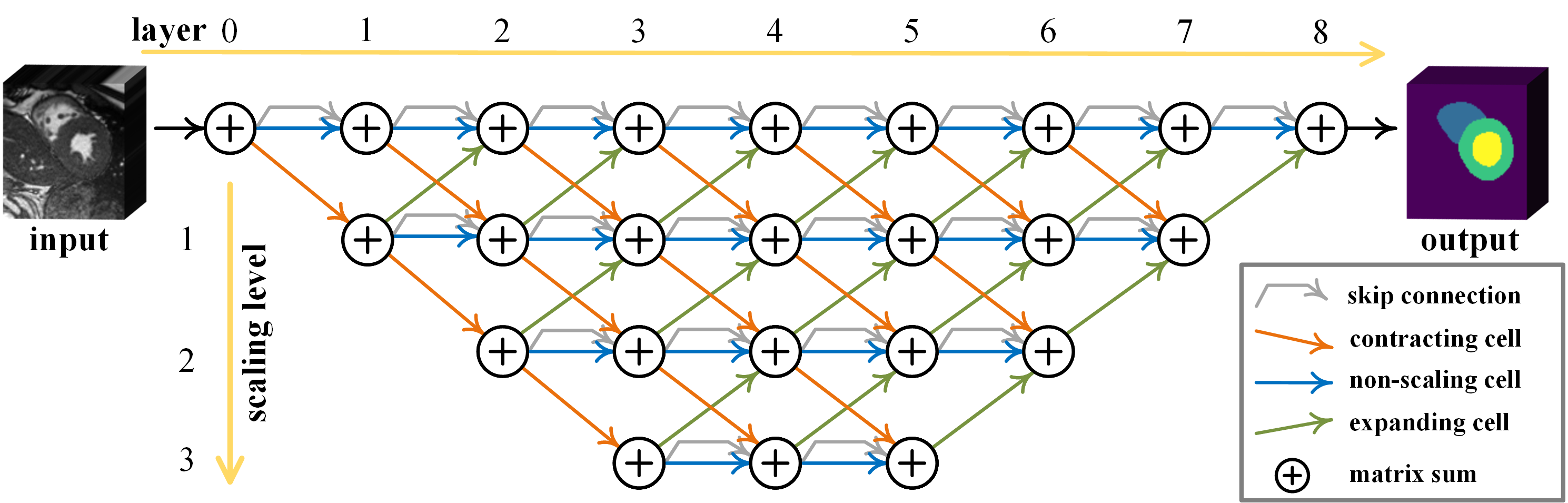}
\caption{Overview of our 3D multi-scale architecture search space for real-time Cine MRI segmentation. The network level search space can fuse features from different scaling level, thus including various existing network designs.}
\label{figure_1}
\end{figure*}

\section{Related Work}

In order to reduce the efforts to design neural architecture manually, researchers have resort to neural architecture search (NAS) techniques, which can automatically search for optimal neural architectures on specific tasks \cite{liu2018darts, cai2018proxylessnas} such as image classifications.
Some works have shown that the neural architectures explored by NAS can indeed outperform hand-crafted ones \cite{weng2019unet, yan2020ms}.
Existing works on NAS mainly focuses on three dimensions: search space, search strategy, and performance estimation strategy \cite{elsken2018neural}.
The search strategy includes random search, Bayesian optimization, evolutionary method, reinforcement learning, and gradient-based methods.
Recently, gradient-based search strategy is becoming more prevailing because of its high speed and less hardware resource \cite{liu2018darts,xu2019pc}.
Usually, in differentiable NAS, an over-parameterized supernet containing multiple paths from the input to the output is constructed. Each path is associated with some architecture weights which are differentiable with respect to the loss function of the supernet.
After the search finishes, the paths with the largest weights are kept to derive the final architecture and other redundant paths are pruned \cite{liu2018darts}.
In \cite{liu2018darts, xu2019pc, cai2018proxylessnas, weng2019unet}, a fixed network backbone is used for searching the repeated cell structure.
In \cite{liu2019auto, yan2020ms}, not only the cell structure but also the network backbone can be explored during a search.
Therefore, with some acceleration methods like partial channel connection \cite{xu2019pc} and binarized path \cite{cai2018proxylessnas}, better accuracy can be achieved.
Most recently, several works have explored hardware-aware NAS \cite{wu2019fbnet,tan2019mnasnet, li2020edd, lu2019neural,cai2018proxylessnas}.
\cite{wu2019fbnet, cai2018proxylessnas, li2020edd} incorporate hardware latency as an additional term into the objective function of supernet.
\cite{tan2019mnasnet} treats latency as a constraint on the objective.

The success of NAS in image classification task has boosted the development of NAS for semantic image segmentation tasks \cite{liu2019auto,chen2018searching}, especially in the field of medical image segmentation \cite{weng2019unet, yan2020ms, kim2019scalable, dong2019neural, mortazi2018automatically}.
For example, \cite{weng2019unet} proposed NAS-Unet which uses U-net as the architecture backbone and parallel search for two types of cell structure (e.g., DownSC and UpSC).
Experimental results show that NAS-Unet has a better performance compared with vanilla U-net and FC-DenseNet \cite{jegou2017one}(a variant U-net).
\cite{yan2020ms} introduces a Multi-Scale NAS framework which is featured with multi-scale search space and multi-scale fusion of features at different scales for medical image segmentation.
A two-step approach is used to decode the final architecture from the relaxed architecture parameters.
However, unlike in NAS for image classification, most existing NAS works for medical image segmentation do not take hardware latency into consideration. Therefore, they can hardly satisfy the real-time constraints needed by the visual assistance in cardiac intervention.

\section{Method}

\subsection{Overview of the neural architecture search structure}

In this section, We will first introduce the network structure and 3D multi-scale neural architecture search space.
Then we will demonstrate how we incorporate the latency estimated as an additional term into the loss function and jointly search for network architecture with high accuracy and low latency.
Finally, we will discuss how the network is optimized and how to decode the discrete architecture once the search finishes.
Figure \ref{figure_1} shows an overview of the 3D multi-scale neural architecture search space, which can be represented as a directed acylic graph (DAG).
Each vertex $v_i$ represents a feature map. Each edge $e_{ij}$ represents an operation between two vertices $v_i$ and $v_j$.
Following the recent cell-based neural architecture search strategy \cite{liu2018darts, xu2019pc, weng2019unet}, the best cell architecture is searched during the neural architecture search phase and then shared across the entire network.
Note that a cell is defined as a network structure associated with each edge in the DAG.

\begin{figure}[t]
\centering
\includegraphics[width=1.0 \columnwidth]{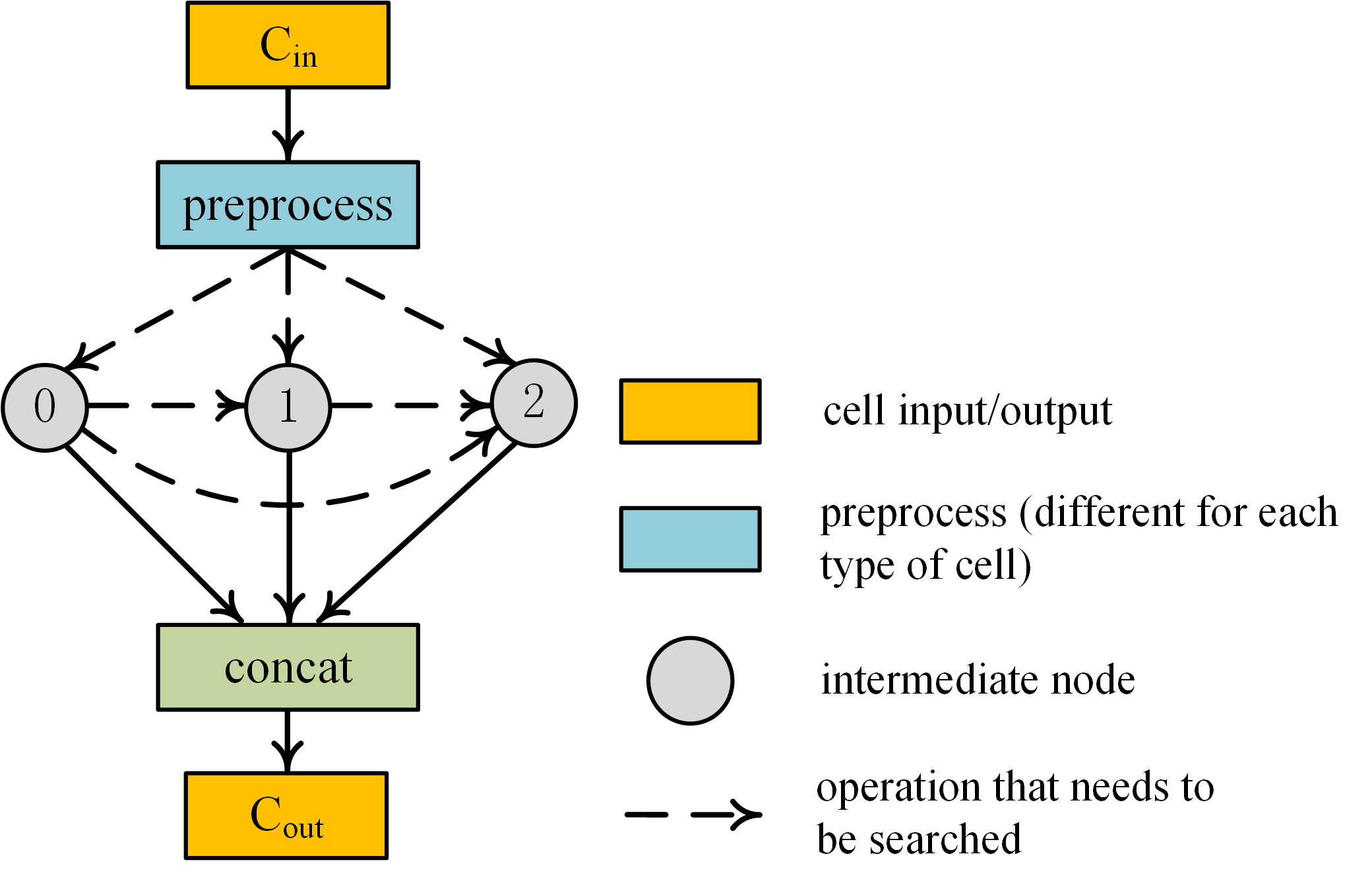}
\caption{An example of cell structure with three intermediate nodes. The structure is shared among non-scaling cell, contracting cell and expanding cell. The difference between three types of cell lies inside the preprocessing step. The dashed lines indicate the primitive operations (e.g., identity operation or dilation convolution operation) that need to be searched.}
\label{figure_2}
\end{figure}

\subsection{Three types of cell structures}

Like existing cell-based NAS works \cite{weng2019unet, liu2018darts}, the structure of the cell is shared by the entire network, different cell structures can be defined according to the functionality of the cell.
In our work, we define three types of cell architectures called non-scaling cell, contracting cell and expanding cell which are associated with the non-scaling edge, the contracting edge, and the expanding edge in the DAG in 
Figure \ref{figure_1} respectively.
The non-scaling cell keeps the scale of the feature map. The contracting cell and the expanding cell are used to down-sample and up-sample the scale of the feature map.

Figure \ref{figure_2} shows the structure of the defined cells.
The structure is shared among three types of cells discussed above.
Each types of cell has different preprocessing block, which contains a scaling operation (e.g., maxpooling or upsampling).
The dashed edges/lines inside the cell represent the primitive operations that need to be searched.
The primitive operations will be discussed in section 3.3.
Suppose there are $m$ intermediate nodes inside a cell. Then the total number of edges that need to be searched is $m(m+1)/2$.
In the end, the outputs of all intermediate nodes are concatenated into a single feature map as the output of the cell.

\subsection{Search space}

\begin{figure}[]
\centering
\includegraphics[width=1.0 \columnwidth]{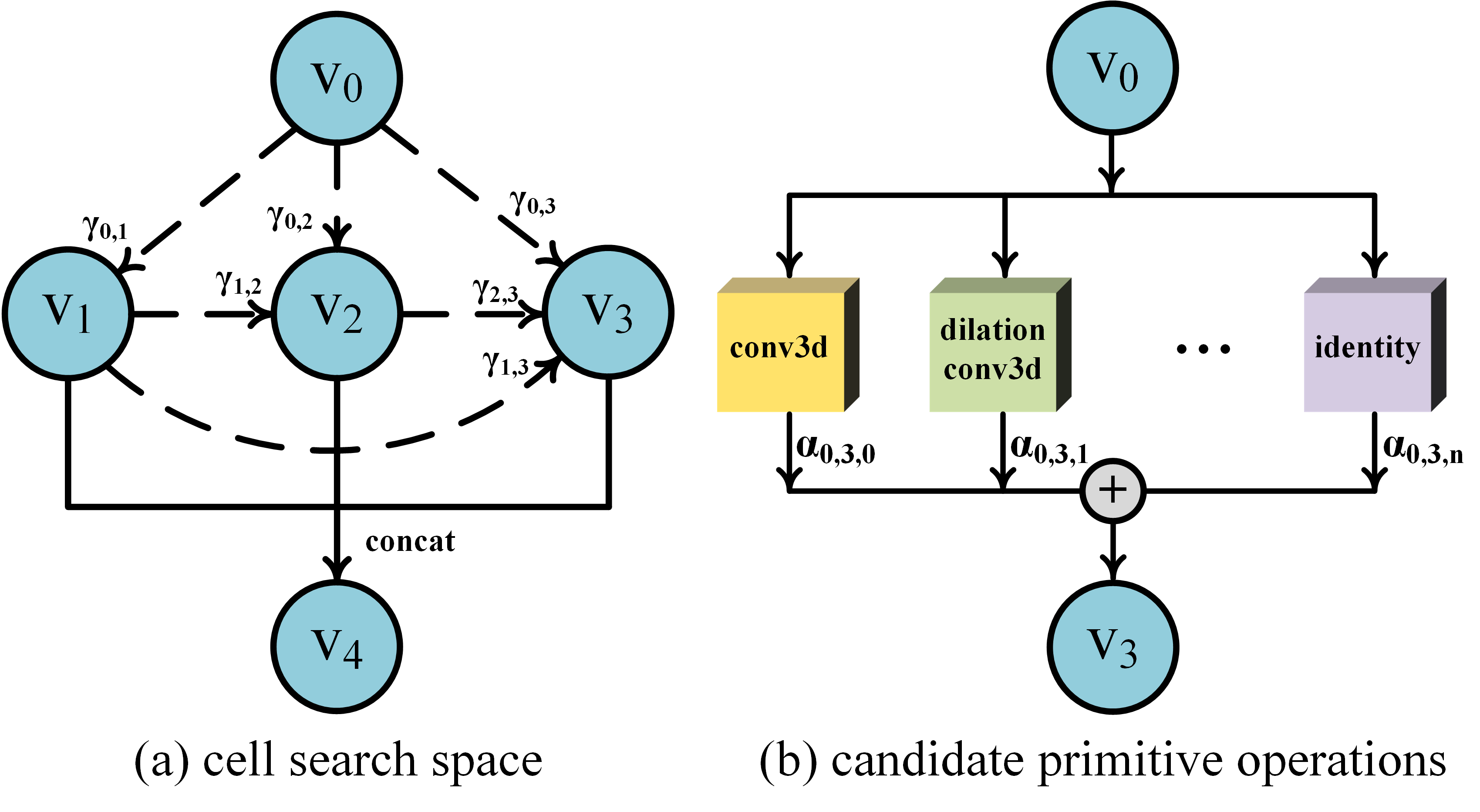}
\caption{(a) Cell search space with three intermediate nodes in the cell. (b) Illustration of over-parameterized mixed operations. Each edge is associated with $N$ candidate primitive operations. $v_0$ is the feature map of the output of the preprocessing block. $v_4$ is the feature map of cell output. $\gamma_{i,j}$ is the architecture parameter associated with the edge from $v_i$ to $v_j$. $\alpha_{i,j,k}$ is another architecture parameter that represents the weight of choosing primitive operation $k$ on that edge.}
\label{figure_3}
\end{figure}

\begin{table}[h]
\centering
\caption{Search space of primitive operations.}
\resizebox{\columnwidth}{!}{
\begin{tabular}{ccc}
\hline
Type & Primitive operations & Parameters \\\hline
1 & conv3d & kernel:3, stride:1, padding:1\\
2 & dilation conv3d & kernel:3, stride:1, padding:2, dilate:2 \\
3 & separable conv3d & kernel:3, stride:1, padding:1 \\
4 & maxpool3d & kernel:3, stride:1, padding:1\\ 
5 & identity & None \\
6 & zero & None \\ \hline
\end{tabular}}
\label{table_1}
\end{table}

The search space of our network contains cell level and architecture level.
At the cell level, we can search for the best primitive operations between intermediate nodes as well as the edge connections among them.
As shown in Figure \ref{figure_3} (a), suppose there are $3$ intermediate nodes in a cell, then the search space in the cell can be represented as a DAG consisting $5$ vertices, with $v_0$ be the output feature map of the preprocessing block and $v_4$ be the output feature map of the cell.
Each edge represents a mixture of primitive operations.
In our work, the search space for the primitive operations is normal 3D convolution, dilation 3D convolution, depth-wise-separable 3D convolution, 3D maxpooling, identity (skip connection) and zero (none operation) as shown in Table \ref{table_1}.
Recent works \cite{liu2018darts,cai2018proxylessnas} showed the feasibility and effectiveness of using continuous relaxation for an over-parameterized network that includes all candidate paths. We use the same strategy for our neural architecture search.
Specifically, it can be seen in Figure \ref{figure_3} (b) that each edge is a mixture of $N$ candidate primitive operations. Let $OP=OP_k$ be one of the primitive operations from the $N$ candidate sets and $X_j$ be the feature map of node $v_j$. 
Then the feature map of $v_j$ can be calculated as:
\begin{equation}
    X_j = \sum_{i<j}{p_{\gamma_{i,j}} \cdot \sum_{k\in N}p_{\alpha_{i,j,k}} \cdot OP_k(X_i)}, j>0.
\label{equation_1}
\end{equation}
where $\alpha_{i,j,k}$ is the weight of the corresponding primitive operation $k$ on the edge from $v_i$ to $v_j$, $p_{\alpha_{i,j,k}}$ is the probability of choosing this operation. $\gamma_{i,j}$ is the weight parameter of edge from $v_i$ to $v_j$. $p_{\gamma_{i,j}}$ represents the probability of this edge.
During search, $p_{\alpha_{i,j,k}}$ is computed by using softmax on the weights of all candidate operations $\alpha_{i,j}$. $\gamma_{i,j}$ is computed by applying softmax on weights of all edges to coming to node $v_i$.
After neural architecture search, we can prune the redundant operations and paths by choosing $\alpha_{i,j,k}$ and $\gamma_{i,j}$ with the highest probability.
\begin{equation}
    p_{\alpha_{i,j,k}} = e^{\alpha_{i,j,k}} / \sum_{k \in N}(e^{\alpha_{i,j,k}}),
\end{equation}
\begin{equation}
    p_{\gamma_{i,j}} = e^{\gamma_{i,j}} / \sum_{i<j}(e^{\gamma_{i,j}}).
\end{equation}

\begin{figure}[t]
\centering
\includegraphics[width=1.0 \columnwidth]{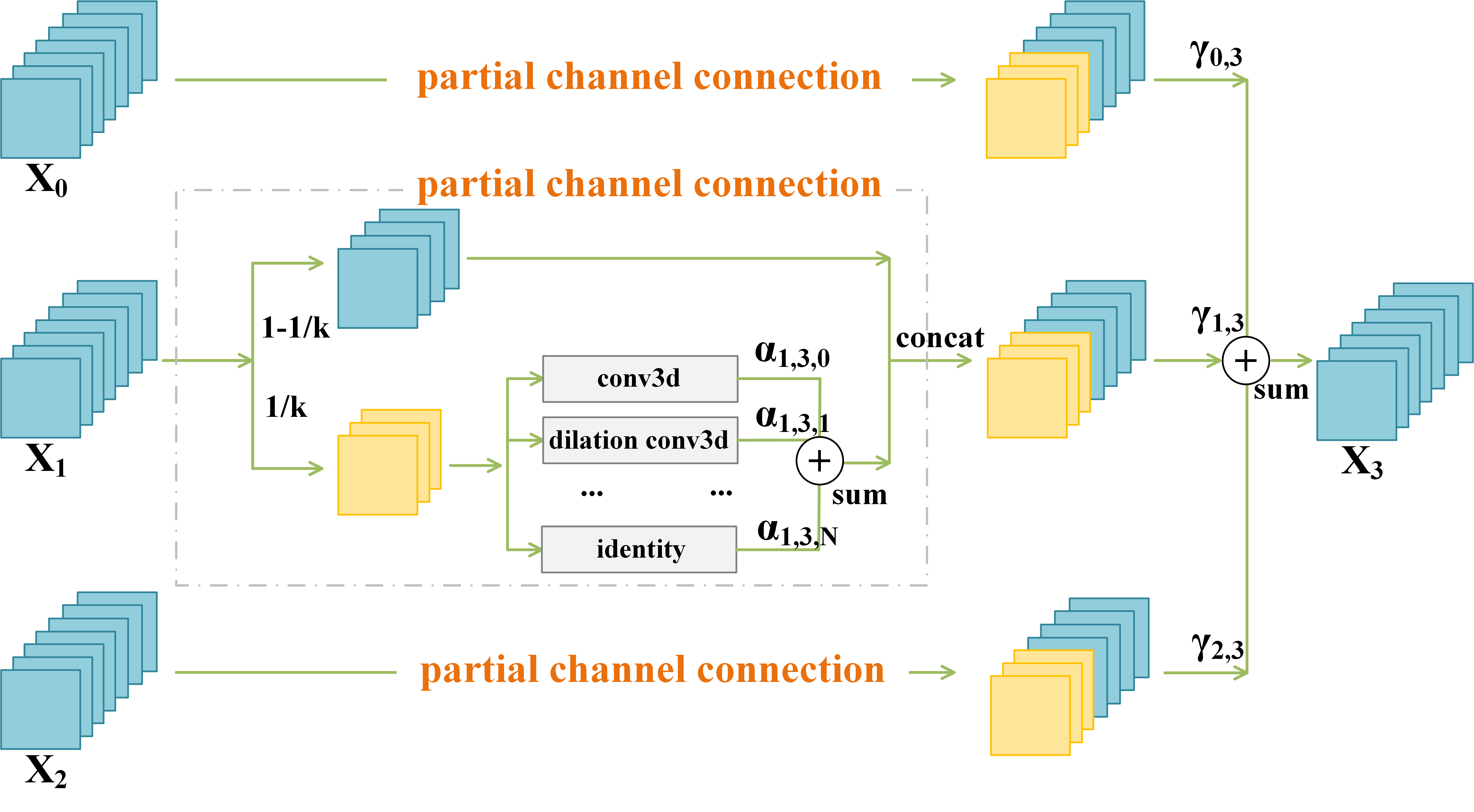}
\caption{Calculation of vertex $v_3$ in a cell with partial channel connection.}
\label{figure_4}
\end{figure}

Inspired by \cite{xu2019pc}, we further apply a partial channel connection strategy in the search scheme to reduce the memory cost during a search.
Specifically, we use a hyper-parameter $k$ to split the channels of the feature map into two parts, $1/k$ part and $1-1/k$ part.
Let $1/k$ part do the mixed primitive operations and the remaining part stay unchanged. Finally, the results of these two parts are concatenated to form the resulting feature.
Therefore, we can rewrite equation \ref{equation_1} to:
\begin{equation}
    X_j = \sum_{i<j}{\gamma_{i,j} \cdot (\sum_{k\in N}\alpha_{i,j,k} \cdot OP_k(k_{i,j}\times X_j) + (1-k_{i,j})\times X_j)}, j>0.
\end{equation}
in which $k_{i,j}$ is a sampling mask that samples $1/k$ portion of channels from tensor $X_i$.
In Figure \ref{figure_4}, we use the calculation of feature map $X_3$ in a cell to demonstrate the mechanism of partial channel connection.
The idea of partial channel connection is to let part of the channels go through the operations and the rest part stay constant, thus reducing the computation.

As for the network level search, unlike Darts \cite{liu2018darts} and Nas-unet \cite{weng2019unet} which fix the network backbone during the search, 
our framework has a multi-scale search space that can identify various backbones, such as UNet, ResUnet, FCN, etc. 
In particular, in the network level, we hope to search for the connection between vertices in Figure \ref{figure_1} (not vertices in the cell) to find the optimal path (or sub-network) within the entire network search space.
For each path from the input to output in the DAG, there is only one edge going into a vertex and one edge coming out of it. Neural architecture search aims to identify such path that can produce the best accuracy and latency.
Since we are using the continuous relaxation search strategy, each edge is assigned with a weight parameter $\beta$, indicating the sampling probability of this connection in the graph.
Let $X_s^l$ be the feature map in layer $l$. Scaling level $s$, $s \in \{0,1,2,3\}$. $Ep(*)$, $Ns(*)$ and $Ct(*)$ denote the expanding, non-scaling and contracting operations respectively. 
$\beta _{i,j}^{l-1}$ denotes the weight parameter of edge from $X_i^{l-1}$ to $X_j^{l}$, $\beta _{i}^{l-1}$ is the weight parameter of the skip connection from $X_i^{l-1}$ to $X_i^{l}$.
Then the feature map $X_s^l$ can be calculated as:
\begin{equation}
\begin{split}
    X_s^l = &p_{\beta_{s+1,s}^{l-1}} \cdot Ep(X_{s+1}^{l-1})+p_{\beta_{s-1,s}^{l-1}} \cdot Ct(X_{s-1}^{l-1}) \\
    & + p_{\beta_{s,s}^{l-1}} \cdot Ns(X_{s}^{l-1}) + p_{\beta_{s}^{l-1}} \cdot X_s^{l-1}.
\end{split}
\end{equation}

Similar to cell structure parameters $\alpha$ and $\gamma$, softmax operation is also used to normalize real parameter $\beta$ to get the corresponding probability $p_\beta$.
Note that $\beta$ is normalized based on the weights of all the edges that come out of a vertex. Some vertices may not contain certain edges.
For example, the vertices in the first layer do not have expanding edges that come out of them.


\subsection{Hardware-aware loss function}

Most existing NAS methods focus on reducing FLOPs. However, FLOPs may not always reflect the actual latency performance of a network \cite{wu2019fbnet, cai2018proxylessnas}.
In order to simultaneously optimize model accuracy and efficiency, we apply the following loss function during search
\begin{equation}
min: \mathcal{L}(w,\alpha, \beta, \gamma) = CE(w, \alpha, \beta, \gamma) + \lambda \cdot LAT(\alpha, \beta, \gamma).
\end{equation}
where $w$ is the network weight parameters. $\alpha$, $\beta$ and $\gamma$ are architecture parameters.
The first term $CE(w,\alpha, \beta, \gamma)$ is the cross-entropy loss for accuracy optimization. The second term $LAT(\alpha, \beta, \gamma)$ is the hardware-aware latency term aiming to optimize the latency performance of the searched network on the target device (GPU in our case).
$\lambda$ is a hyper-parameter that controls the magnitude of the latency term.
$\lambda$ is set to 0.0001 in all our experiments.
Similar to FBNet \cite{wu2019fbnet}, we use a latency lookup table to estimate the overall latency of the network based on the runtime of each primitive operation.

Prior hardware-aware NAS methods \cite{wu2019fbnet, cai2018proxylessnas, li2020edd} usually use the fixed backbone during a search.
Therefore, their latency can be estimated by taking the summation of all operators under the assumption that the runtime of each operator is independent of others.
This is true if the network search space only contains a single backbone structure.
However, in our multi-scale framework, the data dependency is more complicated. The information on one vertex can choose different paths to go and those operations on different paths may run simultaneously using GPU parallelism.
Therefore, simply sum the latency of all operations in the network may not reflect the actual latency of the searched network.
To address this issue, we introduce a path-wise latency estimation method, in which the expected latency of the $n$ longest paths are estimated and used as the expected latency of the searched network.
Intuitively, only the $n$ longest paths are kept in the final searched network and our goal is to minimize the latency of these paths. 
Like \cite{yan2020ms}, $n$ is a hyperparameter that determines the trade-off between model complexity and efficiency, which will be discussed in section 3.5. 

\textbf{Expected latency at the cell level:} Similar to previous work of latency estimation, we use a differentiable paradigm to solve the problem.
In particular, the expected latency inside a cell is calculated by summing up the latency of preprocessing operation and all mixed operations among intermediate nodes.
The expected latency of the mixed operation between vertices in a cell is calculated as
\begin{equation}
\begin{split}
    \mathop{{}\mathbb{E}}[lat_{s,s+1}^{l}[{MixedOP_{i,j}}]] = \sum_{k\in N}{GS({\alpha_{i,j,k}}|\alpha_{i,j})\cdot F(OP_k)}.
\end{split}
\label{equation_7}
\end{equation}
in which $lat_{s,s+1}^{l}[{MixedOP_{i,j}}]$ represents the latency of the mixed operation from $v_i$ to $v_j$ in a cell from layer $l$ to $l+1$ and scaling level $s$ to $s+1$.
$GS(\cdot)$ is the Gumbel-Softmax sampling rule to make the architecture parameters differentiable to the latency loss term.
$\alpha_{i,j}$ are the weight parameters of the primitive operations from $v_i$ to $v_j$. $\alpha_{i,j,k}$ is the weight parameter of the $k\-$th parameter from the candidate primitive operations. 
The indices of vertices in a cell are illustrated in Figure \ref{figure_3}.
$N$ is the total number of candidate primitive operations.
$F(\cdot)$ is the latency estimating function that returns the latency of some operation $OP_k$.
With the latency of mixed operation defined, we can then compute the expected latency of the cell as
\begin{equation}
\begin{split}
    &\mathop{{}\mathbb{E}}[lat_{s,s+1}^{l}] = lat_{s,s+1}^{l}[\textit{preprocessing}] \\  & +  \sum_j^M \sum_i^j{GS({\gamma_{j,t}}|\gamma_{j})}\cdot \mathop{{}\mathbb{E}}[lat_{s,s+1}^{l}[{MixedOP_{i,j}}]].
\end{split}
\label{equation_8}
\end{equation}
where $lat_{s,s+1}^{l}$ is the latency of a cell from layer $l$ to $l+1$ and scaling level $s$ to $s+1$.
$\gamma_{j}$ are the weight parameters of edges coming into vertex $v_j$, $\gamma_{j,t}$ is the weight from one of these edges.

\textbf{Expected latency at the architecture level:}
Based on the expected cell latency estimated above, we define the latency of a path to be the summation of the cell latency of each layer in the path. This is based on the observation that the input feature of layer $l$ is only dependent on the features of the previous layer $l-1$. 
Specifically,
\begin{equation}
\begin{split}
    &\mathop{{}\mathbb{E}}[lat[path_{n}]] = \sum_{l \in L}{GS({\beta_{s,t}^{l}}|\beta_{s}^l)\mathop{{}\mathbb{E}}[lat_{s,t}^{l}]}       \quad  \beta_{s,t}^{l} \in path_{n}.
\end{split}
\label{equation_9}
\end{equation}
$\mathop{{}\mathbb{E}}[lat[path_{n}]]$ is the expected latency of top $n$-th longest path in the searched network. We will discuss the definition of path length and how to find the top $n$ longest path in Section 3.5.
$L$ is the number of layers. $lat_{s,t}^{l}$ is the latency of all possible cells from layer $l$ to $l+1$ and scale $s$ to $t$. $\beta_{s,t}^l$ is the corresponding weight parameter.
Then the expected latency of the entire network is simply the summation of latency of all top $n$ longest paths.
It can be seen from Equation \ref{equation_7} to Equation \ref{equation_9} that the expected latency term is differentiable respect to architecture parameter $\alpha$, $\beta$, and $\gamma$. 
Therefore, by minimizing the latency loss, we are searching for the optimal architecture parameters with the best latency performance.

\subsection{Searching strategy and decoding}

Similar to \cite{liu2018darts, liu2019auto}, we formulate the architecture search as a continuous optimization problem.
Specifically, we adopt the first-order approximation in \cite{liu2018darts} for searching.
The training set is split into two parts $train_{weight}$ and $train_{arch}$ for weight parameter optimization and architecture parameter optimization respectively.
The training contains two phases. For the first few epochs, the architecture parameters are frozen and only network weight parameter $w$ is updated by $\bigtriangledown_w\mathcal{L}_{train_{weight}}(w,\alpha, \beta, \gamma)$. Then, for the rest training epochs, network weight parameter $w$ and architecture parameters $\alpha, \beta, \gamma$ are updated in turn by $\bigtriangledown_w\mathcal{L}_{train_{weight}}(w,\alpha, \beta, \gamma)$ and $\bigtriangledown_{\alpha, \beta, \gamma}\mathcal{L}_{train_{arch}}(w,\alpha, \beta, \gamma)$.

Once the search finishes, we still need to decode the architecture parameters to final discrete architecture.
At the cell level, we take the argmax on $\alpha$ to choose the most likely primitive operation. We also take the argmax on $\gamma$ to find the edge connection between vertices in a cell.
Note that each vertex in a cell represents a feature map, and a latter vertex may connect to all of its previous vertices. Examples can be seen in Figure \ref{figure_3}. $\gamma$ is the architecture parameter that controls these connections.
In the final cell architecture, one vertex can only connect with one of its previous vertices.

\begin{table*}[h]
\tabcolsep 14pt
\renewcommand\arraystretch{1.3}
\caption{Comparison between baseline results and our proposed hardware-aware NAS on Dice score, latency (LT) and throughput (TP). ``D" and ``IF" denote the depth and initial filter channel of U-net. $n$ represents the number of fusion paths in MS-NAS. Note that in order to satisfy the real-time constraints, latency should be below 50 ms to avoid visual lags, and throughput should be above MRI reconstruction rate of 22 FPS.}
\begin{tabular}{lcccccc}
\hline
\multirow{2}{*}{Methods} & \multicolumn{4}{c}{Dice score} & \multirow{2}{*}{TP (FPS)} & \multirow{2}{*}{LT (ms)} \\ \cline{2-5}
& LV   & RV   & MYO   & Average  &&\\ \hline
U-net 2D (D5, IF=64)     &  0.911$\pm$.026 & 0.865$\pm$.036   &  0.761$\pm$.039 & 0.846$\pm$.025 &  16.1&62\\
U-net 2D (D3, IF=16)     &  0.767$\pm$.026    & 0.564$\pm$.071     &   0.738$\pm$.045    & 0.690$\pm$.036    &    \textbf{33.3}  &    \textbf{30}                 \\
U-net 3D (D4, IF=32)     &   0.905$\pm$.027   &   0.855$\pm$.039   &  0.830$\pm$.044     &     0.863$\pm$.032     &         6.4            &              157       \\
U-net 3D (D3, IF=16)     &  0.886$\pm$.029    &  0.805$\pm$.059    &  0.790$\pm$.034     &  0.827$\pm$.033       &        17.2            &          58           \\
MS-NAS 3D ($n$=4)             &  \textbf{0.911$\pm$.030}    &  \textbf{0.873$\pm$.039}    &    \textbf{0.858$\pm$.033}   &   0.880$\pm$.036       &        7.2             &   139                  \\ \hline
our method ($n$=2) &   0.911$\pm$.029   &  0.865$\pm$.041    & 0.850$\pm$.033   &  0.875$\pm$.034     &     22.2    & 45 \\
our method ($n$=1)  &  0.905$\pm$.030   &  0.856$\pm$.038    &   0.842$\pm$.033    &    0.867$\pm$.034      &     25.6    & 39 \\
\hline
\end{tabular}
\label{table_2}
\end{table*}

Unlike the cell level decoding, at the architecture level, we cannot simply apply the same strategy on $\beta$ to obtain the final discrete architecture path.
This is because ideally, we can only have one edge in each layer in a single path, and argmaxing on edges coming out of different vertices in the same layer makes no sense as they are not related.
Note that the sum of weights of edges coming out of a vertex is always 1 and the probability of choosing the current edge is conditioned on the probability of previous edges.
Therefore, we let each network vertex sample independently. Then the length of a path can be described as the product of all the edge weights in the path
\begin{equation}
    length(path_n) = \prod_{l\in L}{p_{\beta^{l}}^{(n)}}.
\end{equation}
where $p_{\beta^l}^{(n)}$ is the softmax probability of the edge's parameter $\beta$ in layer $l$ of path $n$.
Note that for different path $n$, $p_{\beta^l}^{(n)}$ may be different.
Dynamic programming is used for finding the $n$-th longest path in the DAG.

In order to further improve segmentation performance, we use the same strategy as in \cite{yan2020ms}. Instead of using the longest path as the final searched architecture, we fuse the top-$n$ longest paths in the DAG.
Intuitively, larger $n$ represents a bigger network model and more computation resource, but higher model complexity and usually better segmentation accuracy.
Therefore, $n$ is a hyperparameter that can be used to explore tradeoff between accuracy and hardware efficiency.
Experiment results will show that a small level of path fusion (n=2) can improve latency while maintaining competitive accuracy.

\section{Experimental Results}

This section demonstrates the experimental results of the proposed neural architecture search framework on the ACDC MICCAI 2017 challenge dataset.
Results demonstrate that our proposed method can reduce the inference latency up to 3.5 times, compared to the state-of-the-art NAS framework \cite{yan2020ms}, thus satisfying real-time constraints.

\subsection{Dataset}
We evaluate our neural architecture search framework using the ACDC MICCAI 2017 challenge dataset \cite{bernard2018deep} with additional labeling done by experienced radiologists \cite{wang2019msu}.
The task is to segment right ventricle (RV), myocardium (MYO) and left ventricle (LV) from 3D cardiac MRI cine for both end-diastolic and end-systolic phases instances.
The dataset contains 150 exams from different patients with 100 for training and 50 for testing.
Dice score and 5-fold cross-validation are used for evaluating the segmentation accuracy.
we evaluate the test data by submitting the segmentation results of ED and ES instants to ACDC online evaluation platform \cite{acdc}.

\subsection{Experiment setup}

During the searching phase, we split the training set of ACDC by 1:1 into two parts: $train_{weight}$ and $train_{arch}$, each with 50 cases.
$train_{weight}$ is used for network weight training and $train_{arch}$ is for architecture parameter training.
After a search finishes, the searched architecture is then trained from scratch on the entire training set following the 5 fold cross-validation.
We compare our work with the state-of-the-art NAS framework MS-NAS \cite{yan2020ms} for medical image segmentation.
We modify MS-NAS to a 3D version to handle 3D MRI images.
We also implement traditional 3D U-Net with different configurations (D4, IF32) and (D3, IF16), 2D U-net with different configurations (D5, IF64) and (D3, IF16). For example, (D4, IF32) represents the depth of the network is 4, the initial filter number is 32.
2D U-net is based on the implementation of \cite{wang2019msu}.

\begin{figure*}[t]
\centering
\includegraphics[width=2.0 \columnwidth]{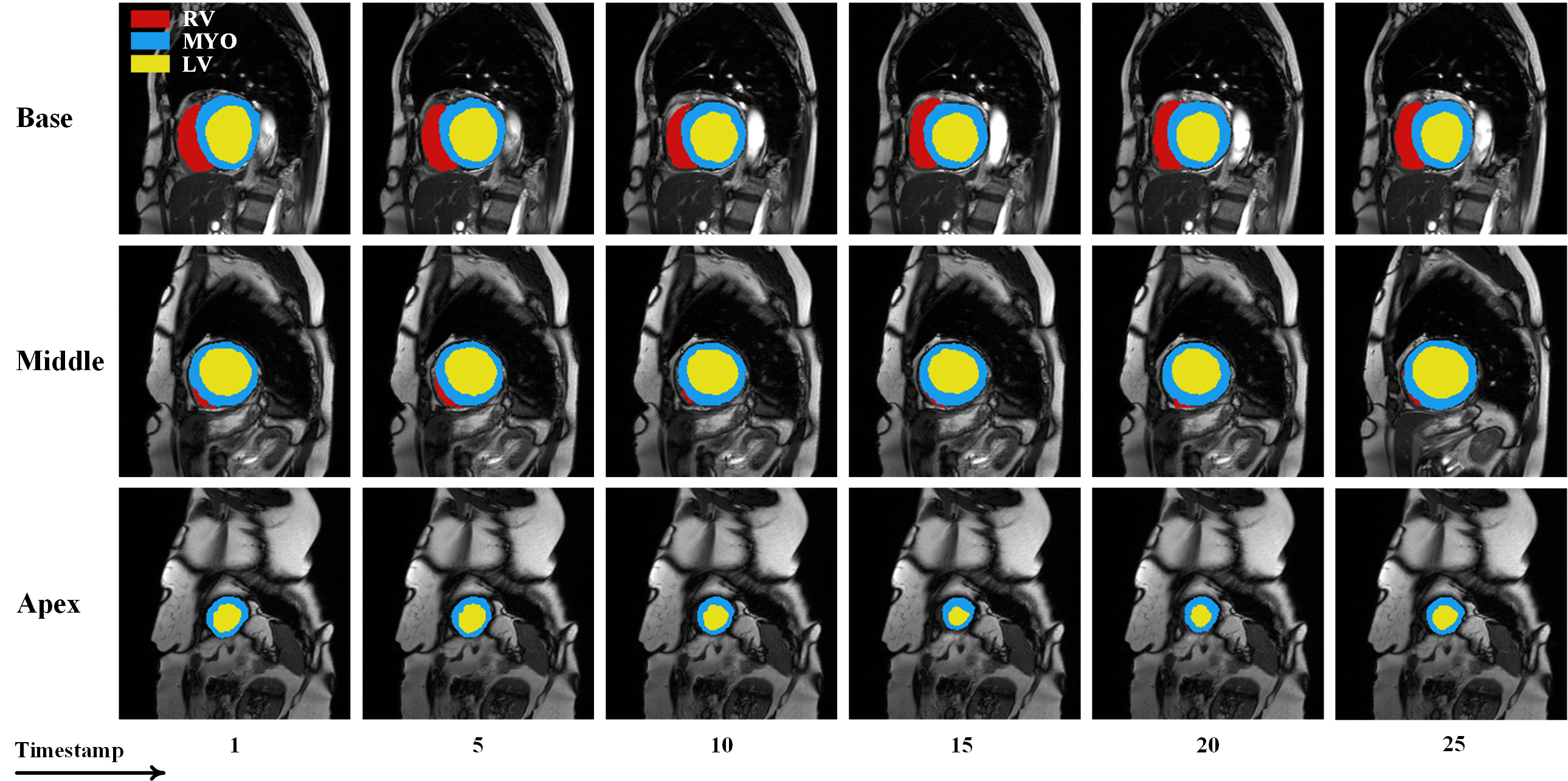}
\caption{Qualitative segmentation results of our method ($n=2$) on the test data. The rows represent the slices at the base, the middle and the apex of LV. The columns represent the segmentation of different timestamps. The red, blue and yellow indicate right ventricle (RV), myocardium (MYO) and left ventricle (LV), respectively.}
\label{figure_6}
\end{figure*}

In our 3D multi-scale NAS supernet, the number of layers is set to 8. The number of intermediate nodes in a cell is 3.
We set $k=4$ for partial channel connection.
The 3D cine MRI images are resized and cropped to 20$\times$128$\times$112.
3D cine MRI images are preprocessed using the same strategy as \cite{isensee2018nnu} with the same data augmentation for a fair comparison.
We optimize the network weight in the first 30 epochs. For the rest 30 epochs, network weight and network architecture parameters are optimized alternately.
All these methods are fully trained with the same hyperparameters.

The evaluated networks and NAS frameworks are implemented using Pytorch.
All experiments run on a machine with 16 cores of Intel Xeon E5-2620 v4 CPU, 256G memory, and four NVIDIA GeForce GTX 1080 GPUs.

\begin{figure}[t]
\centering
\includegraphics[width=1.0 \columnwidth]{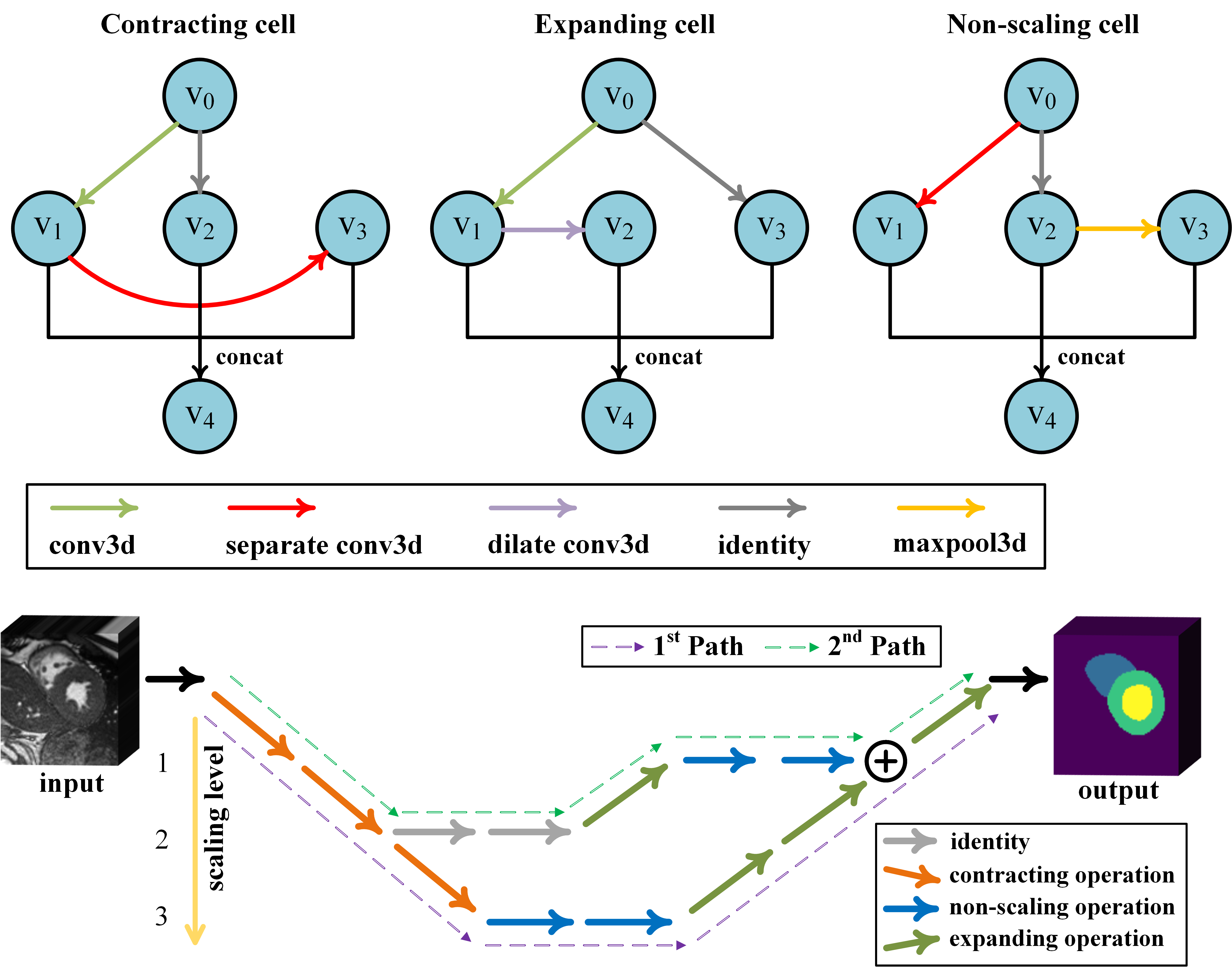}
\caption{Illustration of searched cell structure (including contracting cell, expanding cell and non-scaling cell), and searched network architecture. Best viewed in color.}
\label{figure_5}
\end{figure}

\subsection{Results}

\noindent\textbf{A. Comparison with the existing models.}

Table \ref{table_2} shows the comparison among 2D U-net, 3D U-net, NAS-Unet, MS-NAS and our proposed NAS framework on the ACDC 3D cardiac cine MRI dataset.
From the table, we can see that MS-NAS 3D can achieve the best Dice score on all of the three anatomy structures.
However, it cannot satisfy the timing constraint.
On the other hand, our proposed approach can guarantee the timing constraint to be met (i.e., 3.5 times improvement on latency), while achieving competitive accuracy against MS-NAS, with at most of 0.8\% accuracy loss or even without accuracy loss.
Note that $n$ represents the number of fusion paths in the searched network. Larger $n$ means larger network, usually with better accuracy.

We have several detailed observations from this table: (1) For all competitors, only U-net 2D with the depth of 3 and the initial filter channels of 16 (denoted as U-net 2D-3-16) can satisfy the throughput constraint of ($\ge$22FPS), while our method can guarantee that the throughput meets the constraint.
This is because i) for 3D medical images, the network FLOPs are usually very large. We have to carefully design the network structure in order to fulfill the real-time constraint; ii) MS-NAS uses much larger operators in the cell search space (e.g., a modified 3$\times$3 depth-wise-separable convolution which involves 4 convolution layers); iii) MS-NAS does not consider latency while searching.
(2) The Dice score of U-net 2D-3-16 for LV, RV, and MYO are merely 0.767, 0.564, and 0.738, while ours can improve it to 0.911, 0.865, and 0.850, respectively.
(3) Even compared with the results of U-net 3D-5-64 and U-net 3D-4-32 that cannot satisfy the real-time constraint, our method can achieve better accuracy.
(4) With $n=1$, we can achieve better latency at the cost of lower Dice score. Specifically, compared to $n=2$, $n=1$ can achieve 6ms latency improvement. However, the average Dice score drops by about 0.9$\%$.
This demonstrates that the proposed method can provide 
freedom for designers to make a better tradeoff on accuracy and hardware efficiency.

All the above results verify the importance of conducting hardware-aware NAS for real-time 3D cardiac cine MRI segmentation.


\noindent\textbf{B. Visualization of the resultant architecture.}

Figure \ref{figure_5} presents the searched cell structure (top) and network structure (bottom).
In the identified network structure, two paths with the longest length are fused to form the final architecture.
The search is a trade-off between accuracy and latency. It can be seen in the non-scaling cell that the network prefers weight-free operations (e.g, identity and maxpooling). This is because the non-scaling cell is less important than other cells (e.g., contracting cell) in segmentation networks.

There is an interesting observation on the second-longest path. It is shallower than the first path and upsampling is not done until the last layer.
One potential reason is that NAS is trying to guarantee the timing performance of the searched network. Such a path is hard to identify manually, which is why we need NAS to help explore network structures.

\noindent\textbf{C. Visualization of the segmentation results.}

Finally, Figure \ref{figure_6} shows the visualization of some segmentation examples at different timestamps in a cine MRI. 
We can observe that for various positions (e.g., base and apex) and different timestamps, our method can always accurately segment the targets.
It can be seen that the RV segmentation results are not as accurate as LV and MYO. This is because in some slices the RV is very small and the boundary is unclear, which makes it hard to segment.

\section{Conclusion}

In this paper, we propose a hardware-aware NAS framework for real-time 3D cardiac cine MRI segmentation.
We explicitly incorporate the target GPU hardware latency into the network objective function and jointly optimize both the segmentation accuracy and hardware performance.
We formulate our problem as an over-parameterized differentiable structure so that traditional gradient descent optimization can be used.
We introduce a 3D multi-scale search space which can fuse feature with different tensor sizes to achieve higher segmentation accuracy.
Experimental results on ACDC MICCAI 2017 dataset show that our searched architecture can satisfy real-time throughput and latency requirements to visually guide cardiac interventions, while still maintaining high accuracy. 

\bibliographystyle{ACM-Reference-Format}
\bibliography{reference.bib}

\end{document}